# Bio-authentication based secure transmission system using steganography

Najme Zehra,
IGIT, GGSIP University
Delhi, India
nznaqvi.igit@gmail.com

Mansi Sharma
IGIT, GGSIP University
Delhi, India
freedom.mansi@gmail.com

Somya Ahuja
IGIT, GGSIP University
Delhi, India
thesomya@gmail.com

Shubha Bansal
IGIT, GGSIP University
Delhi, India
theshubha@gmail.com

*Abstract*— Biometrics deals with identity verification of an individual by using certain physiological or behavioral features associated with a person. Biometric identification systems using fingerprints patterns are called AFIS (Automatic Fingerprint Identification System). In this paper a composite method for Fingerprint recognition is considered using a combination of Fast Fourier Transform (FFT) and Sobel Filters for improvement of a poor quality fingerprint image. Steganography hides messages inside other messages in such a way that an "adversary" would not even know a secret message were present. The objective of our paper is to make a bio-secure system. In this paper bio–authentication has been implemented in terms of finger print recognition and the second part of the paper is an interactive steganographic system hides the user's data by two options- creating a songs list or hiding the data in an image.

*Keywords*--Fingerprint,minutiae,Listega, steganography,LSB

## I. INTRODUCTION

**B**iometrics consists of methods for uniquely identifying humans based upon one or more intrinsic physical or behavioral traits. In information technology, biometrics is used as a form of identifying access management and access control [1]. A biometric system is a pattern recognition system that operates by getting biometric data from a person, extracting a feature set from the acquired data, and comparing this feature set against the template set in the database. Fingerprint recognition is one of the oldest methods of biometric identification. It is popular because of the inherent ease in acquisition, the numerous sources (ten fingers) immigration. Steganography's goal in general is to hide data well enough that unintended recipients do not suspect the steganographic medium of containing hidden information. Contemporary approaches are often classified based on the steganographic cover type into image, audio, graph, or text. A steganography approach must be capable of passing both computer and human examination. This can be achieved by the list based and image based steganography techniques. We have used the LSB technique for image based steganography.

## II. FINGERPRINT RECOGNITION

A fingerprint consists of ridges, which are lines across fingerprints, and valleys, which are spaces between ridges. The pattern of the ridges and valleys is unique for each individual.

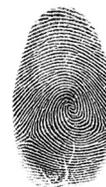

Fig. 1. A fingerprint showing valleys and ridges

There are three approaches to fingerprint matching techniques:
*1)Correlation-based matching:* In this method, two fingerprint images are superimposed and the correlation between corresponding pixels is calculated for different alignments.

*2)Minutiae-based matching:* In this method, minutiae are extracted from the two fingerprints and stored as sets of points in the two- dimensional plane. After that, alignment between the template and the input minutiae sets are found that results in the maximum number of minutiae pairings.

*3)Pattern-based matching:* This algorithm tries to do matching based on the global features (arch, whorl, and loop) of a whole fingerprint image with a previously stored template. For this the images are aligned in the same orientation. To do this, the algorithm selects a central point in the fingerprint image and centers on that. The template contains the type, size, and orientation of patterns within the aligned fingerprint image. The candidate fingerprint image is graphically compared with the template to determine the degree to which they match. The proposed system is classified into various modules and sub-modules as given in Figure 2. It has two major modules: Minutiae Extraction and Minutiae Matching.

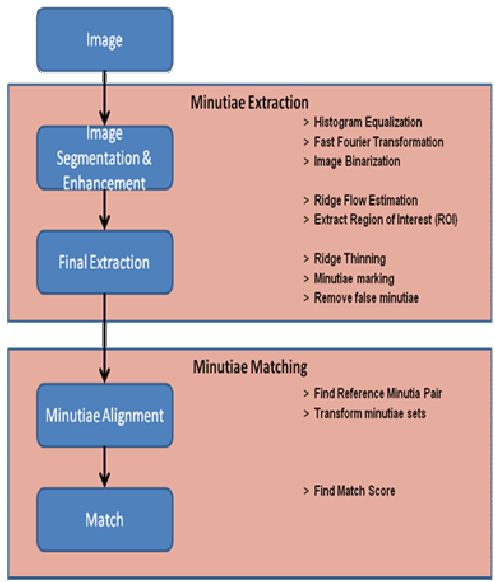

Fig. 2.Modules of the proposed system

## A. Minutiae Extraction

The Minutiae extraction process consists of image enhancement, image segmentation and final Minutiae extraction.

The first step in the minutiae extraction stage is **Fingerprint Image enhancement**. The goal of an enhancement algorithm is to improve the clarity of the ridge structures in the recoverable regions and mark the unrecoverable regions as too noisy for further processing. These enhancement methods can increase the contrast between ridges and furrows and for join the false broken points of ridges due to insufficient amount of ink.

In our paper we have implemented three techniques: Histogram Equalization, Fast Fourier Transformation and Image Binarization.

*1) Histogram equalization:* It is a technique for improving the global contrast of an image by adjusting the intensity distribution on a histogram. This allows areas of lower local contrast to gain a higher contrast without affecting the global contrast. Histogram equalization spreads out the most frequent intensity values. The original histogram of a fingerprint image has the bimodal type (Figure 3.1(a)), the histogram after the histogram equalization occupies all the range from 0 to 255 and the visualization effect is enhanced (Figure 3.1(b)).

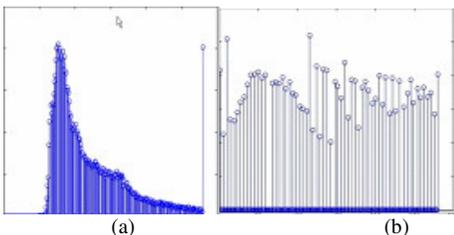

Figure 3.1(a) Original histogram, (b) Histogram after equalization

The result of the histogram equalization is shown in figure 3.2.

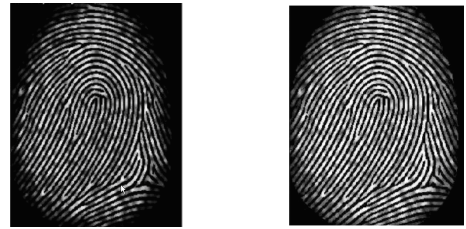

(a)        (b)

Figure 3.2(a) Original Image, (b) Enhanced Image after histogram equalization

*2)Fast Fourier Transformation:* Fourier Transform is an image processing tool which is used to decompose an image into its sine and cosine components. The output of the transformation represents the image in the frequency domain, while the input image is the spatial domain equivalent.

In this method the image is divided into small processing blocks (32 x 32 pixels) and Fourier transform is performed according to equation:

$$F(u,v) = \sum_{x=0}^{M-1}\sum_{y=0}^{N-1} f(x,y) \times \exp\left\{-j2\pi \times \left(\frac{ux}{M} + \frac{vy}{N}\right)\right\} \quad (1)$$

for u = 0, 1, 2, ..., 31 and v = 0, 1, 2, ..., 31.

In order to enhance a specific block by its dominant frequencies, we multiply the FFT of the block by its magnitude a set of times, where the magnitude of the original FFT = abs (F (u, v)) = |F (u, v)|.

So we get the enhanced block according to the equation:

$$g(x,y) = F^{-1}\left\{F(u,v) \times |F(u,v)|^{k}\right\} \quad (2)$$

where F⁻¹(F (u, v)) is given by:

$$f(x,y) = \frac{1}{MN}\sum_{x=0}^{M-1}\sum_{y=0}^{N-1} F(u,v) \times \exp\left\{j2\pi \times \left(\frac{ux}{M} + \frac{vy}{N}\right)\right\} \quad (3)$$

For x = 0, 1, 2 ...31 and y = 0, 1, 2 ...31.

The k in formula (2) is an experimentally determined constant, which we choose k=0.45 to calculate. A high value of k improves the appearance of the ridges by filling up small holes in ridges, but too high value of k can result in false joining of ridges which might lead to a termination become a bifurcation.

Figure 4 presents the image after FFT enhancement.

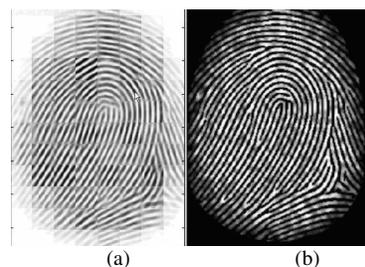

(a)        (b)

Figure 4(a) Enhanced Image after FFT, (b) Image before FFT

The enhanced image after FFT has the improvements as some falsely broken points on ridges get connected and some spurious connections between ridges get removed.

*3) Image Binarization:* After this, the fingerprint image is binarized using the locally adaptive threshold method like the Sobel Filter. Fingerprint Image Binarization is done to transform the 8-bit Gray fingerprint image to a 1-bit image with 0-value for ridges and 1-value for furrows. The Sobel Filter transforms a pixel value to 1 if the value is larger than the mean intensity value of the current block (16x16) to which the pixel belongs.

The Sobel filter is used in image processing, particularly within edge detection algorithms. Technically, it is a discrete differentiation operator, calculating an approximation of the gradient of the image intensity function. At each point in the image, the result of the Sobel operator is either the corresponding gradient vector or the norm of this vector. The Sobel operator is based on convolving the image with a small, separable, and integer valued filter in horizontal and vertical direction and is therefore relatively inexpensive in terms of computations. On the other hand, the gradient approximation which it produces is relatively crude, in particular for high frequency variations in the image. Mathematically, the Sobel operator uses two 3×3 kernels which are convolved with the original image to calculate approximations of the derivatives - one for horizontal changes, and one for vertical. If A is the source image, and Gx and Gy are images which at each point contain the horizontal and vertical derivative approximations, the computations are as follows:

$$G_y = \begin{bmatrix} +1 & +2 & +1 \\ 0 & 0 & 0 \\ -1 & -2 & -1 \end{bmatrix} * A \quad \text{and} \quad G_x = \begin{bmatrix} +1 & 0 & -1 \\ +2 & 0 & -2 \\ +1 & 0 & -1 \end{bmatrix} * A$$

where * here denotes the 2-dimensional convolution operation. The x-coordinate is here defined as increasing in the "right"-direction, and the y-coordinate is defined as increasing in the "down"-direction. At each point in the image, the resulting gradient approximations can be combined to give the gradient magnitude, using:

$$G = \sqrt{G_x^2 + G_y^2}$$

Using this information, we can also calculate the gradient's direction:

$$\Theta = \arctan\left(\frac{G_y}{G_x}\right)$$

where, for example, $\Theta$ is 0 for a vertical edge which is darker on the left side.

After image enhancement the next step is fingerprint image segmentation. Segmentation is the breaking of an image into two components i.e. the *foreground* and the *background*. The foreground originates from the contact of a fingertip with the sensor. The noisy area at the borders of the image is called the background. The task of the fingerprint segmentation algorithm is to decide which part of the image belongs to the foreground and which part to the background [1]. In general, only a Region of Interest (ROI) is useful to be recognized for each fingerprint image. The image area without effective ridges and furrows is first discarded since it only holds background information. Then the bound of the remaining effective area is determined since the minutiae in the bound region are confusing with those spurious minutiae that are generated when the ridges are out of the sensor.

To extract the region of interest, two steps are followed: Block direction estimation and ROI extraction by Morphological methods. The Block direction estimation involves two steps:

a.. Estimate the block direction for each block of the fingerprint image with WxW in size (W is 16 pixels by default).
The algorithm is:
I. Calculate the gradient values along x-direction (gx) and y-direction (gy) for each pixel of the block. Two Sobel filters are used to fulfill the task.

II. For each block, use following formula to get the Least Square approximation of the block direction.

$$\tan \beta = \frac{\sum_{i=1}^{w} \sum_{j=1}^{w} g_x g_y}{\sum_{i=1}^{w} \sum_{j=1}^{w} (g_x^2 - g_y^2)}$$

for all the pixels in each block. The formula is easy to understand by regarding gradient values along x-direction and y-direction as cosine value and sine value. So the tangent value of the block direction is estimated nearly the same as the way illustrated by the following formula.

$$\tan \beta = \frac{2 \sin \beta \cos \beta}{(\cos 2\beta - \sin 2\beta)}$$

b. After finished with the estimation of each block direction, those blocks without significant information on ridges and furrows are discarded based on the following formulas:

$$E = \frac{2\{(g_x g_y) + (g_x^2 - g_y^2)\}}{W * W * (g_x^2 + g_y^2)}$$

For each block, if its certainty level E is below a threshold, then the block is regarded as background block. For Region of Interest (ROI) extraction Two Morphological operations called

'OPEN' and 'CLOSE' are adopted. The 'OPEN' operation can expand images and remove peaks introduced by background noise. The 'CLOSE' operation can shrink images and eliminate small cavities.

*Final Minutiae Extraction:*

After enhancement of the image and segmentation of the required area, minutiae extraction requires four operations: Ridge Thinning, Minutiae Marking, False Minutiae Removal and Minutiae Representation.

*1)Ridge Thinning:*It is the process of removing the redundant pixels of ridges till the ridges are just one pixel wide. This is done using the MATLAB's built in morphological thinning function.

bwmorph(binaryImage,'thin',Inf)

The thinned image is then filtered, again using MATLAB's three morphological functions to remove some H breaks, isolated points and spikes (Figure 5).

bwmorph(binaryImage, 'hbreak', k)
bwmorph(binaryImage, 'clean', k)
bwmorph(binaryImage, 'spur', k)

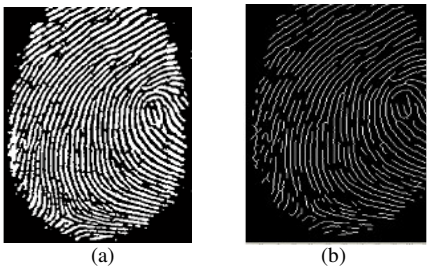

(a)  (b)

Figure 5(a) Image before, (b) Image after thinning

*Minutiae Marking:*

Minutiae marking are performed using templates for each 3 x 3 pixel window as follows. If the central pixel is 1 and has exactly 3 one-value neighbors, then the central pixel is a ridge branch (Figure 6.1).

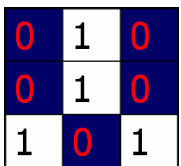

Figure 6.1

If the central pixel is 1 and has only 1 one-value neighbor, then the central pixel is a ridge ending (Figure 6.2).

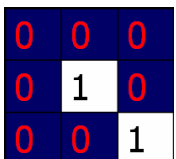

Figure 6.2

There is one case where a general branch may be triple counted (Figure 6.3). Suppose if both the uppermost pixel with value 1 and the rightmost pixel with value 1 have another neighbor outside the 3x3 window due to some left over spikes, so the two pixels will be marked as branches too, but actually only one branch is located in the small region.

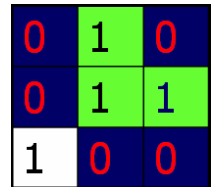

Figure 6.3

*False Minutiae Removal*

To keep the recognition system consistent the false minutiae must be removed. For the removal we first calculate the inter ridge distance D which is the average distance between two neighboring ridges. For this each row is scanned to calculate the inter ridge distance using the formula:

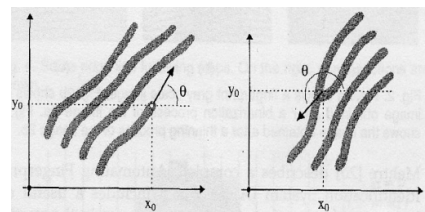

Inter ridge distance = $\dfrac{sum\ all\ pixels\ with\ value\ 1}{row\ length}$

Finally an averaged value over all rows gives D.

After this we will be labeling all thinned ridges in the fingerprint image with a unique ID for further operation using a MATLAB morphological operation BWLABEL.

1) If d (bifurcation, termination) < D & the 2 minutia are in the same ridge then we have to remove both of them.
If d (bifurcation, bifurcation) < D & the 2 minutia are in the same ridge then we have to remove both of them.

2)If d(termination, termination) $\approx$ D & their directions are coincident with only a small angle variation & no other termination is located between the two terminations then we have to remove both of them.

3) If d (termination, termination) < D & the 2 minutia are in the same ridge then we have to remove both of them where d(X, Y) is the distance between 2 minutia points.

*Minutiae Representation*

Finally after extracting the valid minutia points from the fingerprint image they need to be stored in some form of representation common for both ridge ending and bifurcation. So each minutia will be completely characterized by following parameters: x-coordinate, y-coordinate, orientation, and ridge associated with it.

Figure 7:Minutiae Representation

A bifurcation can be decomposed to three terminations each having their own x-y coordinates (pixel adjacent to the bifurcating pixel), orientation and an associated ridge.The orientation of each termination (tx, ty) will be estimated by following method. Track a ridge segment who's starting point is the termination and length is D. Sum up all x-coordinates of points in the ridge segment. Divide above summation with D to get sx. Then get sy using the same way.

Get the direction from:

$$\tan^{-1}\frac{sy-ty}{sx-tx}$$

Results after the minutia extraction stage :

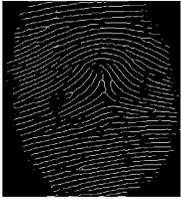

Figure 8.1 Thinned image

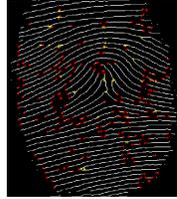

Figure 8.2 Minutiae after marking

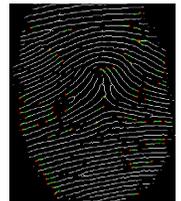

Figure 8.3 Real Minutiae after false removal

*B. Minutiae Matching:*

After successful extraction of the set of minutia points of 2 fingerprint images to be tested, Minutiae Matching is to be performed to check whether they belong to the same person or not. Jain et al [2] provided a method of minutiae matching both at local and global level. On its basis we use a iterative ridge alignment algorithm to align one set of minutiae with respect to other set and then carry-out an elastic match algorithm to count the number of matched minutia pairs. Let $I_1$ & $I_2$ be the two minutiae sets represented as

$$I_1 = \{m_1, m_2 \dots m_M\} \text{ where } m_i = (x_i, y_i, \theta_i)$$

$$I_2 = \{m'_1, m'_2 \dots m'_N\} \text{ where } m'_i = (x'_i, y'_i, \theta'_i)$$

Now we choose one minutia from each set and find out the ridge correlation factor between them. The ridge associated with each minutia will be represented as a series of x-coordinates $(x_1, x_2 \dots x_n)$ of the points on the ridge. A point is sampled per ridge length L starting from the minutia point, where the L is the average inter-ridge length. And n is set to 10 unless the total ridge length is less than 10*L.So the similarity of correlating the two ridges is derived from the following formula which calculates the similarity by using the sets $(x_i..x_n)$ and $(X_i..X_n)$:

$$S = \frac{\sum_{i=0}^{m} x_i X_i}{\left[\sum_{i=0}^{m} x_i^2 X_i^2\right]^{0.5}}$$

where $(x_i..x_n)$ and $(X_i..X_n)$ are the set of x-coordinates for the 2 minutia chosen. And m is minimal value of the n and N value.

If the similarity score is larger than 0.8, then go to step 2, otherwise continue matching the next pair of ridges.

2. Now we have to transform each set according to its own reference minutia and then perform matching in a unified x-y coordinate.

If M ($x$, $y$, $\theta$) be the reference minutia found from step 1(say from $I_1$). For each fingerprint, translate and rotate rest of the minutiae ($xi$, $yi$, $\theta i$) with respect to the M according to the formula:

$$\begin{pmatrix} xi_{new} \\ yi_{new} \\ \theta i_{new} \end{pmatrix} = \begin{bmatrix} \cos\theta & -\sin\theta & 0 \\ \sin\theta & \cos\theta & 0 \\ 0 & 0 & 1 \end{bmatrix} \begin{bmatrix} xi - x \\ yi - y \\ \theta i - \theta \end{bmatrix}$$

The new coordinate system has origin at reference minutia M and the new x-axis is coincident with the direction of minutia M. No scaling effect is taken into account as it is assumed the two fingerprints from the same finger have nearly the same size. Hence, we get the transformed sets of minutiae $I_1$' & $I_2$'.

3. We use an elastic match algorithm to count the matched minutiae pairs by assuming two minutiae having nearly the same position and directions are identical.

Three attributes of the aligned minutiae are used for matching: its distance from the reference minutiae, angle subtended to the reference minutiae, and local direction of the associated ridge. The matching algorithm for the aligned minutiae patterns needs to be elastic, as stated by Xudong Jiang and Wei-Yun Yau [3] since the strict match requiring that all parameters (x, y, q) are the same for two identical minutiae is impossible due to the slight deformations and inexact quantization of minutiae.

The algorithm initiates the matching by first representing the aligned input (template) minutiae as an input (template) minutiae string. The final match ratio for two fingerprints is the number of total matched pair over the number of minutia of the template fingerprint. The score is 100*ratio and ranges from 0 to 100. If the score is larger than a pre-specified threshold, the two fingerprints are from the same finger.

## III. STEGANOGRAPHY

Steganography is a technique with which we can hide data. It is used to transmit a cover media with secret information through public channels of communication between a sender and a receiver avoiding detection from an adversary.

*List-Based Steganography: Listega* manipulates the textual list of data to camouflage both a message and its transmittal. It exploits textual data such as books, music CD's, movie DVD's, etc., to hide messages. The list can be fabricated in order to embed data without raising any suspicion. It encodes a message and then assigns it to legitimate items in order to generate a text-cover in form of a list. It has many benefits such as there is a large demand for the popular data which creates heavy traffic thereby reducing the chances of suspicion. Secondly, Listega does not imply a a particular pattern (noise) that an adversary may look for Moreover; it can be applied to all languages [4].

Listega can be divided in 4 modules:

*1. Domain determination*: Determination of appropriate domain is done to achieve steganographical goal.

*2. Message Encoding*: Encodes a message in required form for the camouflaging process.

*3. Message camouflager*: Generates the list-cover, in which data are embedded by employing the output of step 2.

*4. Communications protocol*: The basic rules about the secret communication are decided.

*Resilience of Listega:* Listega is resilient to the following attacks:

*1. Traffic attack*: The main goal of a traffic attack is to detect unusual association between a sender and receiver. Traffic attacks can be a threat for most of the steganographic techniques regardless of the steganographic cover types used. Traffic analysis is deemed ineffective with Listega. Listega camouflages the transmittal of a hidden message to appear legitimate and thus suspicion is averted. It ensures that the involved parties establish a secret channel by having a well-plotted relationship with each other. Moreover, it imposes the communicating parties to use innocent domains that retain high demand by a lot of people. Such domains create a high volume of traffic that makes it impractical for an adversary to investigate all traffics.

*2. Contrast and comparison attacks:* One of the sources of noise that may alert an adversary is the presence of contradictions in a list-cover. These contradictions may raise suspicion about the existence of a hidden message, especially when they are present in the same document.

Automating the generation of a list-cover through the use of data banks makes the cover resistant to these attacks.

*3. Linguistics attacks*: Listega can pass any linguistic attack by both human and machine examinations. This is because the generated cover is normal text. A statistical attack refers to tracking the profile of the text that has been used. A statistical signature of a text refers to the frequency of words and characters used. As per [5], Listega is resistant to statistical attacks because it is simply opt to use legitimate text that is generated naturally by human. Moreover, the generated textual cover by Listega keeps the same profile as its other peer documents that do not have hidden message. Most alterations introduced by Listega are nonlinguistic and do not produce any flaws (noise), making statistical attacks on list- cover ineffective.

## IV. IMAGE BASED STEGANOGRAPHY

Image Steganography allows two parties to communicate secretly. It allows for copyright protection on digital files using the message as a digital watermark. One of the other main uses for Image Steganography is for the transportation of high-level or top-secret documents between international governments. While Image Steganography has many legitimate uses, it can also be quite harmful. It can be used by hackers to spread viruses and destroy machines, and also by terrorists and other organizations that rely on covert operations to communicate secretly and safely.

Least significant bit (LSB) insertion is a common approach for embedding information in a cover image. As per [6], the least significant bit (in other words, the 8th bit) of some or all of the bytes inside an image is changed to a bit of the secret message. Image steganography suffers from potential of distortion, the significant size limitation of the messages that can be embedded, and the increased vulnerability to detection through digital image processing techniques

## V. FUTURE SCOPE

The work can be used for concealing data by various agencies. It can be extended in future to hide other formats like PDF, or, other image formats. Instead of using a list of songs to hide data, a list of books can be used. The various modules presented in this paper are loosely coupled and hence, can be used anywhere in the industry where secured data transmission is required. The sensitivity to small messages can be improved in the future. Its use can be extended to credit cards where authenticity is the main issue.

## VI. CONCLUSION

The work has combined many methods to build a minutia extractor and a minutia matcher. The combination of multiple methods comes from a wide investigation into research papers. It can be used in areas where efficient bit rate is required. It can be applied to any list of items in any language. The paper gives a real life implementation of a bio-secure system. Some of its limitations are that the fingerprint image file should only be in format TAGGED IMAGE FILE FORMAT (TIFF). Secondly, the text file to be hidden should only be in format TXT.

### REFERENCES

[1] Asker M. Bazen and Sabih H. Gerez, "Segmentation of Fingerprint Images".

[2] A.K. Jain, L Hong and R. Bolle. On-line fingerprint verification. IEEE Transactions on Pattern Analysis and Machine Intelligence, 19(4): 302-314, 1997.

[3] Xudong Jiang, Wei-Yun Yau, "Fingerprint minutiae matching based on local and global structures". Paper appears in Pattern Recognition, 2000, Barcelona, Spain. Proceedings, 15th International Conference, Vol: 2, p:1038-1041

[4] Anderson, R.J. & Petitcolas, F.A.P., "On the limits of steganography", IEEE Journal of selected Areas in Communications, May 1998.

[5] Wang, H & Wang, S, "Cyber warfare: Steganography vs. Steganalysis", Communications of the ACM, 47:10, October 2004.

[6] T. Morkel, J.H.P. Eloff, M.S. Olivier, " An overview of image steganography".

[7] Jayanti Addepalli and Aseem Blackfin, "Processor enhance biometric-Identification Equipment Design".

[8] Wang, H & Wang, S, "Cyber warfare: Steganography vs. Steganalysis", Communications of the ACM, 47:10, October 2004

[9] Abdelrahman Desoky," Listega: list-based steganography methodology",page248.

[10] Intelligent biometric techniques in fingerprint and face recognition By L. C. Jain page 9.

[11] Xudong Jiang, Wei-Yun Yau, "Fingerprint minutiae matching based on local and global structures". Paper appears in Pattern Recognition, 2000, Barcelona, Spain. Proceedings, 15th International Conference, Vol: 2, p:1038-1041.

[12] Mei-Ching Chen, Sos S. Agaian, and C. L. Philip Chen, "Generalized Collage Steganography on Images".

[13] Alfredo C. López, Ricardo R. López, Reinaldo Cruz Queeman, "Fingerprint Recognition".

[14] Chen Zhi-li, Huang Liu-sheng, Yu Zhen-shan, Zhao Xin-xin, Zheng Xue-ling, "Effective Linguistic Steganography Detection".

[15] Anil Jain, Arun Ross, Salil Prabhakar, "Fingerprint matching using minutiae and texture features".

[16] Johnson, N.F. & Jajodia, , S., "Exploring Steganography: : Seeing the Unseen", Computer Journal, February 1998.

[17]http://books.google.co.in/books?hl=en&lr=&id=NxDTSR5Zlz4C&oi=fnd&pg=PA35&dq=how+is++fingerprint+recognition+done+biometrics&ots=3LVMX5ALb&sig=mdwFLSs5Z_19WuWqXVezapcGQXk#v=onepage&q=&=false.

[18] R. Chandamouli, Nasir Memon, "Analysis of LSB based steganographic techniques".

[19] Gualberto Aguilar, Gabriel Sánchez, Karina To scano, Moisés Salinas, Mariko Nakano, Hector Perez, " Fingerprint Recognition".

[20] Mehdi Kharrazi, Husrev T. Sencar, and Nasir Memon, "Image Steganography: Concepts and Practice".

[21]Krenn,R.,"SteganographyandSteganalysis",http://www.krenn.nl/univ/cr- y/steg/article.pdf

[22] Anil K. Jain, Arun Ross and Salil Prabhakar, "An Introduction to Biometric Recognition".

[23] http://en.wikipedia.org/wiki/Biometrics